\pdfoutput=1
\documentclass[]{spie}  %>>> use for US letter paper
\usepackage{amsmath,amsfonts,amssymb}
\usepackage{graphicx}
\usepackage[colorlinks=true, allcolors=blue]{hyperref}
\usepackage{siunitx} % For degree symbol

\title{A Do-it-yourself Spectrograph Kit for Educational Outreach in Optics and Photonics}

\author[a]{Pradip Gatkine} %  Pradip Gatkine, Yiwen Hu, Tiecheng Zhu, Yang Meng, Sylvain Veilleux, Joss Bland-Hawthorn, Mario Dagenais 
\author[b]{Gregorio Zimerman}
\author[a]{Elizabeth Warner}

\affil[a]{Department of Astronomy, University of Maryland, College Park, Maryland 20742, USA}
\affil[b]{Department of Aerospace Engineering, University of Maryland, College Park, Maryland 20742, USA}

\authorinfo{Further author information: (Send correspondence to P. Gatkine)\\E-mail: pgatkine@astro.umd.edu}

\begin{document} 
\maketitle

\begin{abstract}
We designed and built a do-it-yourself spectrograph assembly to demonstrate the concept of spectroscopy, an indispensable tool for exploring the cosmos. This spectrograph is designed for optical band (400-750 nm). It uses a transmission grating to disperse the light and a webcam to measure the spectrum. This spectrograph provides a resolving power ($\lambda/\delta\lambda$) of about 1000. This demonstration involves off-the-shelf materials costing less than \$500, thus making it an easy to build demonstration kit for a school or public setting. The kit is well-suited for performing various science experiments and acquiring hands-on experience for students to learn the concepts such as coherence, spectral orders, resolving power, absorption and emission spectra. All of these concepts are an integral part of modern astronomical observations as well as various other fields in STEM such as biomedical engineering, chemical analysis, food and water quality, etc. This kit is portable and fully modular, making it apt for outreach purposes.
\end{abstract}


\section{INTRODUCTION}
\label{sec:intro}  % \label{} allows reference to this section
The field of astronomy underwent a paradigm shift with the invention of spectroscopy. Some of the very first objects systematically studied with this new technique were the Sun and the stars leading to the first insights into the chemical composition of these astronomical objects that are millions and millions of kilometers away \cite{OpenStax}. It is noteworthy that Helium was first discovered in the emission spectrum of the Solar chromosphere during a solar eclipse \cite{lockyer1869viii}. Over the period of last century, spectroscopy has emerged as an immensely powerful technique to uncover a wealth of not only the chemical, but also the kinematic and cosmological phenomena concerning all the cosmic bodies, from small asteroids in the solar system \cite{vilas2002visible} to massive galaxies billions of light years away \cite{harrison2011history,gatkine2018cgm}. Although astronomical objects became the first subjects of spectroscopic investigation, the technique soon found applications in various modern fields such as biomedical sciences, genetic engineering, petrochemical industry, medical diagnostics, food quality and so on. Spectroscopy has truly become one of most ubiquitous techniques of scientific studies. Therefore, it is imperative for educators to introduce this technique to young students and general masses to capture their curiosity and raise their understanding of the techniques of scientific enquiry.   

%% Adding a figure to show how exoplanets can be studies using spectroscopy
   \begin{figure} [ht]
   \begin{center}
   \begin{tabular}{c} %% tabular useful for creating an array of images 
   \includegraphics[height=3cm]{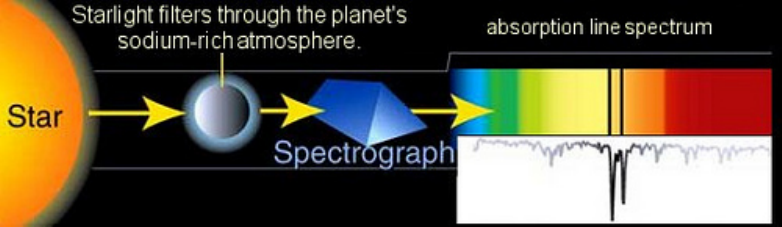}
   \end{tabular}
   \end{center}
   \caption[Picture of the spectrograph] 
%>>>> use \label inside caption to get Fig. number with \ref{}
   { \label{fig:AWG_sim_conventional} 
A schematic to show the study of composition of exoplanet atmospheres using the technique to spectroscopy \cite{NASA}}
   \end{figure}

A promising way to introduce scientific techniques is through an application that can capture the imagination of people with diverse set of interests. One such tantalizing scientific application is - finding exoplanets and searching for potential signs of habitability \cite{grenfell2017review}. With more and more exoplanets being detected and reported in the news, the general population is getting more curious to know about how astronomers find them \cite{summers2017exoplanets}. Therefore, we have designed and built a do-it-yourself spectrograph demonstration and activity with a focus on exoplanet exploration. In addition, we deployed our activity at an event called "Maryland Day" to understand the educational value of such an exhibit.

In this paper, we discuss the design and implementation of our do-it-yourself (DIY) spectrograph and utility of such activities. First, we elaborate upon the design of the spectrograph and the activity. Second, we describe the detailed construction of the spectrograph including the materials. Finally, we discuss the technical capabilities of the spectrograph and its deployment at the 'Maryland Day' event.

  %% Adding AWG-conventional-similarity
   \begin{figure} [ht]
   \begin{center}
   \begin{tabular}{c} %% tabular useful for creating an array of images 
   \includegraphics[height=6cm]{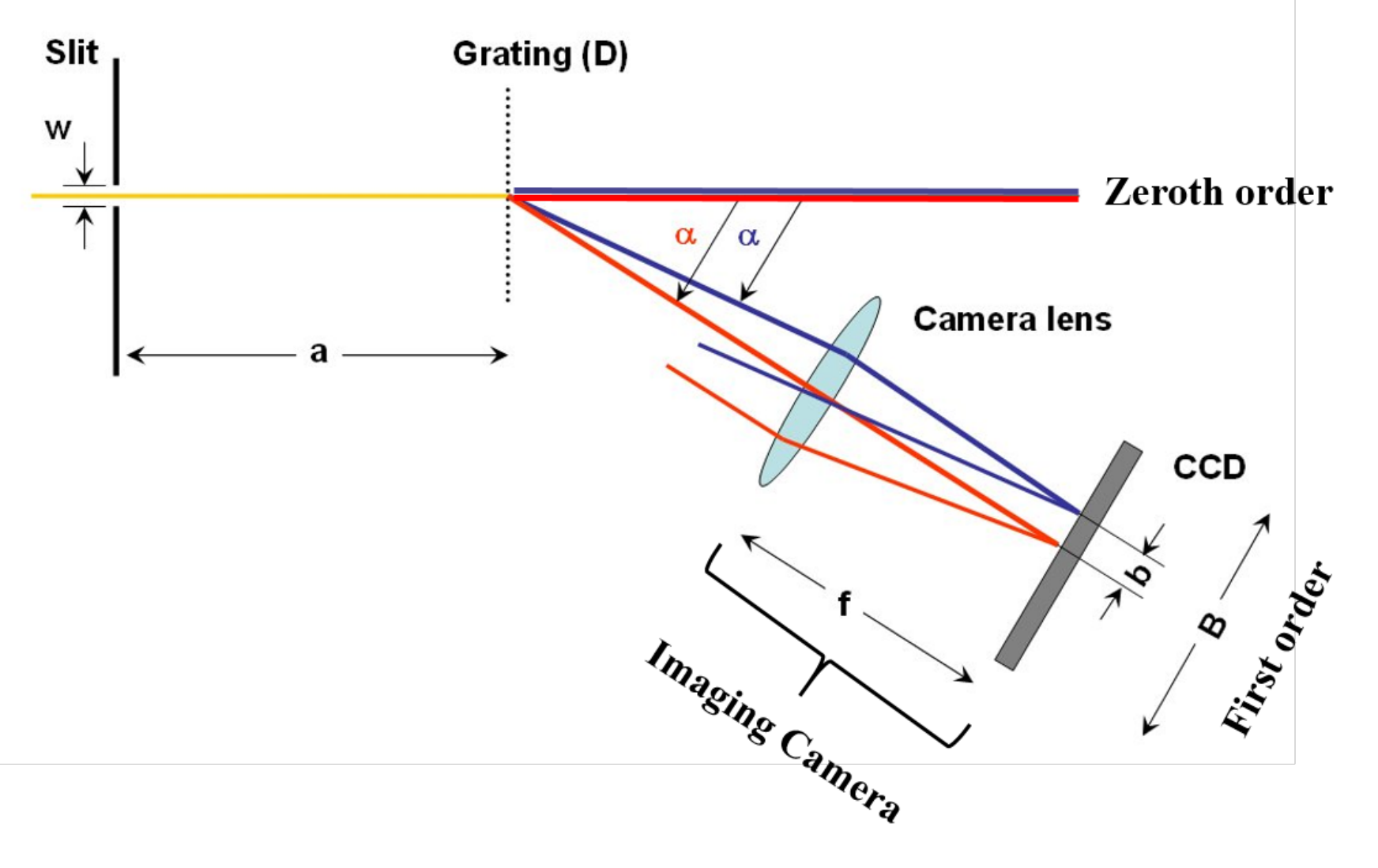}
   \end{tabular}
   \end{center}
   \caption[Picture of the spectrograph] 
%>>>> use \label inside caption to get Fig. number with \ref{}
   { \label{fig:AWG_sim_conventional} 
A simple schematic showing the working of a basic transmission diffraction spectroscopy. The zeroth and first order are shown. Adapted from the Strasbourg project \cite{Strasbourg}
\\}
   \end{figure} 

\section{Design of the Spectrograph}

The focus of the project was to build a DIY spectrograph geared towards a public audience. We chose an iterative design process starting with basic materials then improving material selection after satisfactory testing. Knowing that our audience was the general public, we needed a simple design that could easily demonstrate the capabilities of spectrometers and its working principle, particularly in the search of habitable exoplanets. A spectrometer can be built in numerous ways. The key decision for any spectrometer design is the choice of the dispersing element. There are several possible options including: diffraction gratings (reflective or transmissive), prisms, Echelle gratings \cite{harrison1949production}, holographic gratings \cite{noda1974geometric}, and some advanced on-chip spectrometers \cite{gatkine2016development, gatkine2017arrayed,gatkine2018towards}. We selected use of diffraction gratings since they are the most widely used kind of dispersive element in a variety of applications. Most of the other dispersive elements are some variant of the basic principle of diffraction gratings. Among diffraction gratings, we chose transmission gratings since it is easier to imagine and explain a transmission grating, and it is convenient to design. A typical transmission diffraction grating is shown in Figure 2 with its key elements: the source, slit, grating, and camera. The spectrometer design is dictated by the constraints and desired specifications. These are elicited below.\\

Since this spectrometer is for demonstration and educational purposes, it requires a flexible design to accommodate a suite of experiments in one setup. Therefore, we select the components that provide the necessary flexibility in the setup. In addition, a certain modularity and eas of assembly is desired in the setup to incorporate add-ons for facilitating additional experiments. This, along with the budget constraints dictated the selection of the key elements of the spectrometer:

\noindent
{\bf 1. An adjustable slit:} With an adjustable slit-width, we can showcase the effect of slit-width on the coherence of the light beam and thereby, the spectral resolution. Also, an adjustable width helps to attain the optimal point of the trade-off between the resolving power and the signal to noise ratio. 

\noindent
{\bf 2. A remote controlled light source:} We selected a light source that allows changing the color and brightness using a remote control. This light source is USB powered, thus it can be used with any standard power source. This source offers a selection of 12 colors that is achieved by various combinations of 8 LEDs in the bulb. An add-on feature is the variety of lighting modes which include strobe, flash, and fade which are useful for simulating a transient source as well as adding an entertainment value during basic demonstrations.   

\noindent
{\bf 3. Optical breadboard:} We use an optical breadboard to mount all the optical components in order to add modularity to the design. We can add additional elements on the breadboard such as polarizer, or a sample holder for performing additional experiments. It is made out of aluminum, thus adding the functionality while keeping the system light-weight.\\  

\noindent
{\bf Optical design:} The choice of spectral order is dependent on the requirement of resolving power and the constraint of size of the spectrograph. This can be seen in the schematic of a diffraction grating spectrograph shown in Fig. 2. The choice of the grating was coupled with the requirement of the resolving power (1000). Therefore, we selected a high resolution grating and decided to use the first spectral order, thus keeping the optical system compact and portable. A webcam was selected as the imaging camera because of several advantages such as: USB operation, continuous streaming of the video feed, full wavelength coverage in the optical band, small form factor, easy to mount and adjustable inclination angle to fully sample the first spectral order. With the finalized choice of the components, the dimensions of the setup were tested with a makeshift cardboard box assembly and a CAD for the wooden box was made. This is a two-chamber box with one side holding the light source and the other side housing the optical elements. The details are shown in Fig. 3, 4, and 5. The components we selected for the spectrograph are summarized in Table 1.\\

%% Adding AWG-conventional-similarity
   \begin{figure} [ht]
   \begin{center}
   \begin{tabular}{c} %% tabular useful for creating an array of images 
   \includegraphics[height=4cm]{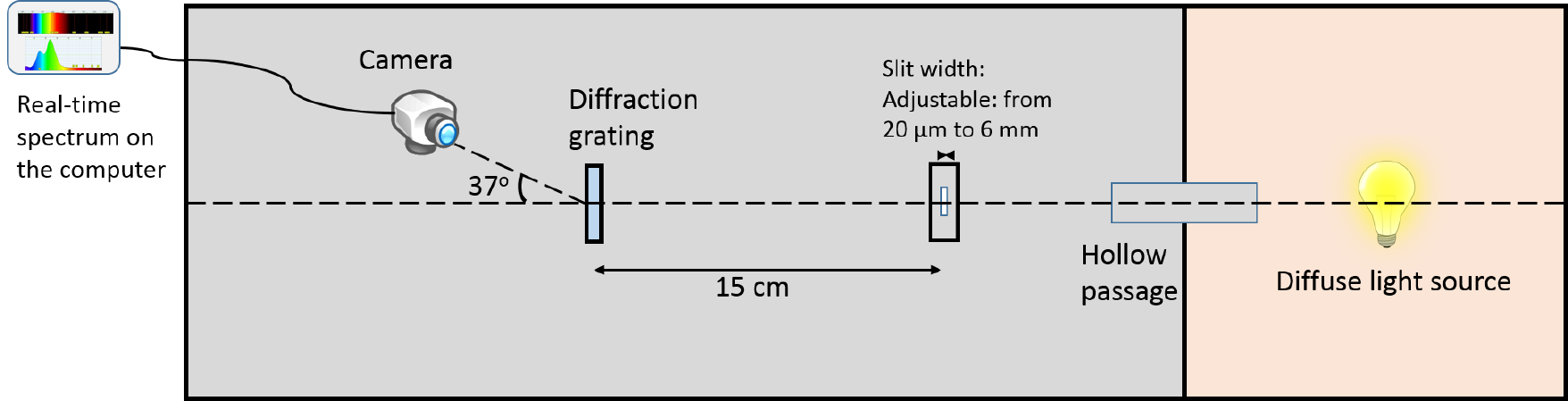}
   \end{tabular}
   \end{center}
   \caption[Picture of the spectrograph] 
%>>>> use \label inside caption to get Fig. number with \ref{}
   { \label{fig:AWG_sim_conventional} 
The framework of the DIY spectrometer presented here including key elements: the source, adjustable slit, diffraction grating, camera and a computer that processes the video feed and generates a spectrum of the source.}
   \end{figure} 

\noindent
{\bf The enclosure:}
The wooden enclosure of the spectrometer is shown in Fig. 4. The enclosure is composed of two chambers separated by partition with a passage to allow the source light to pass through. The left chamber with a transparent acrylic top houses the light source. The transparent top is selected to ensure that the source can be monitored while the experiment is being performed. The right chamber with wooden top (i.e. opaque) houses the optical elements and the camera. In order to minimize the background illumination, the chamber housing the optical components needs to be light-tight so that only the source light enters the chamber. In addition, we paint this chamber in black absorbing paint to ensure that any stray reflected light gets absorbed and doesn't contribute to the background. This helps in achieving the optimal contrast level, which is important for obtaining high-quality spectra.\\

\begin{table}[ht]
\caption{List of components for the spectrometer} 
\label{tab:Component_list}
\begin{center}       
\begin{tabular}{|l|l|} 
\hline
\rule[-1ex]{0pt}{3.5ex} \textbf{Components} & \textbf{Specifications} \\
\hline
\rule[-1ex]{0pt}{3.5ex}  1. Adjustable slit  & Thorlabs VA100/M, range: 20$\mu m$ to 6 mm \\
\hline
\rule[-1ex]{0pt}{3.5ex}  2a. Visible band transmission grating & Thorlabs GT25-12 \\
&  groves: 1200/mm, grove angle: \ang{37}, 25 mm $\times$ 25 mm\\
\hline
\rule[-1ex]{0pt}{3.5ex}  2b. Grating mount  & Thorlabs KM100S - Kinematic Mount for 1" \\
\hline

\rule[-1ex]{0pt}{3.5ex}  3. Imaging Camera & Logitech C920, pixels: 1920 x 1080 pixels, field of view: \ang{78}\\
\hline
\rule[-1ex]{0pt}{3.5ex}  4. Optical breadboard  & Newport SA2\\
& Size: 12" x 4" x 0.5", 1/4-20 holes on a 1" grid\\
\hline
\rule[-1ex]{0pt}{3.5ex}  5. Remote controlled light source  & Sunjack Camplight, 340 lumens, beam angle: 270 degrees \\ 
& (8 LEDs of different colors)\\
\hline
\rule[-1ex]{0pt}{3.5ex}  6a. Calibration Laser: Red   & $\lambda:$ 645 nm, Power: 5mW\\
\hline
\rule[-1ex]{0pt}{3.5ex}  6b. Calibration Laser: Green   & $\lambda:$ 532 nm, Power: 5mW \\
\hline

\end{tabular}
\end{center}
\end{table}

%% Adding AWG-conventional-similarity
   \begin{figure} [ht]
   \begin{center}
   \begin{tabular}{c} %% tabular useful for creating an array of images 
   \includegraphics[height=6cm]{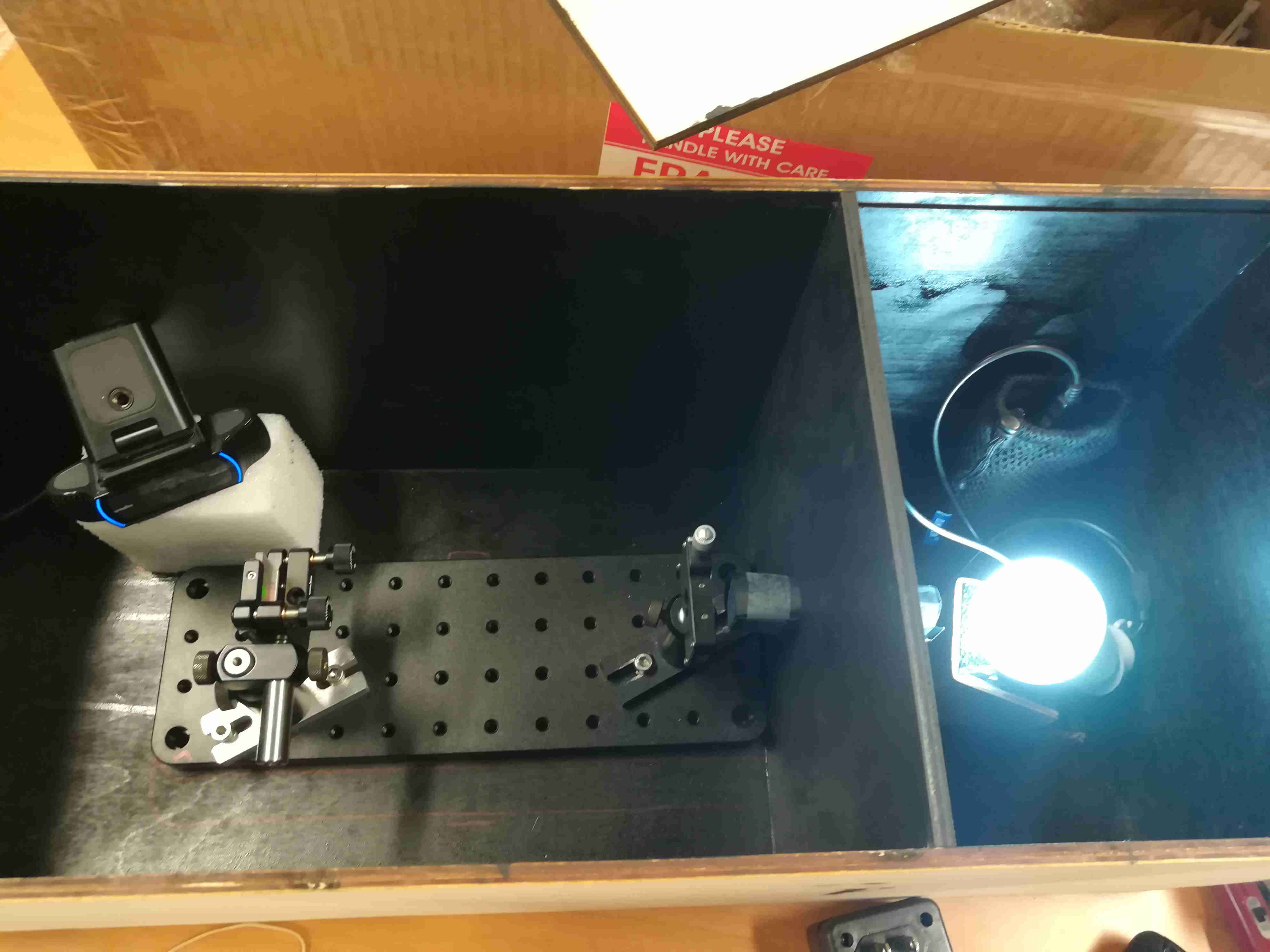}
   \end{tabular}
   \end{center}
   \caption[Picture of the spectrograph] 
%>>>> use \label inside caption to get Fig. number with \ref{}
   { \label{fig:AWG_sim_conventional} 
A picture of the spectrometer showing (from left to right) the source, hollow passage, adjustable slit on the optical breadboard, the transmission grating mounted on the optical breadboard and the webcam at roughly \ang{40} from the optical axis. The chamber containing the source is covered with a transparent acrylic slider (not shown) and the optics chamber is covered on the top by a wooden slider to make it light tight.}
   \end{figure}

%% Adding AWG-conventional-similarity
   \begin{figure} [ht]
   \begin{center}
   \begin{tabular}{c} %% tabular useful for creating an array of images 
   \includegraphics[height=11cm]{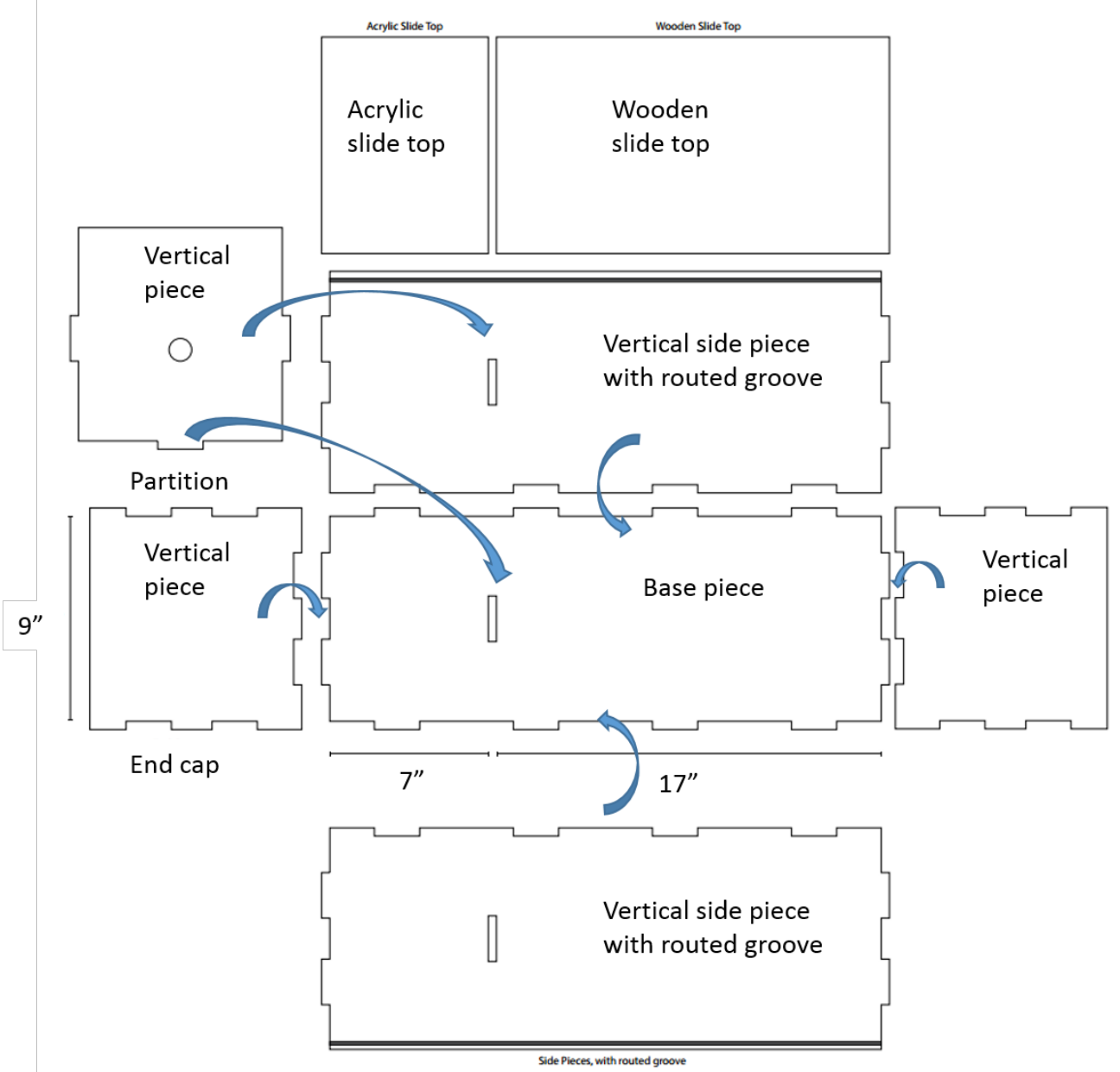}
   \end{tabular}
   \end{center}
   \caption[Picture of the spectrograph] 
%>>>> use \label inside caption to get Fig. number with \ref{}
   { \label{fig:AWG_sim_conventional} 
The CAD of the spectrometer enclosure. There are two chambers in the enclosure: a. Left chamber open to environment, b. Right chamber housing the optics which is light tight and painted black from inside to minimize the background illumination. A vertical plate separates the two chambers with a passage along the principal axis of the optics to allow the light from the source to the optics.\\}
   \end{figure}

\noindent
{\bf The light source:}
We use a remote controlled, USB powered light source to enable wireless control of the source color as well as brightness. It also enables us to explain a important astronomical concept of pulsating stars. At the same time, this features adds a visual/entertainment appeal through the strobe-mode and fading mode operations of the light source.

\noindent
{\bf Image Processing:}
The video stream from the webcam is fed to a computer program called Theremino \cite{Theremino} which is designed to do real-time image processing to transform the image of the spectrum into a plot as a function of wavelength. With the sharp, focused image and the high pixel density of the camera, we obtain a resolving of nearly 1000 with each pixel corresponding to a $\delta \lambda$ of roughly 0.6 nm. The Theremino software also allows calibration of the spectrum using user-defined set points. We use off-the-shelf red (645 nm) and green (532 nm) lasers as our calibration sources. For wavelength calibration, we first shine the red laser from above the light source chamber directly pointed towards the light bulb and turn on the green LED in the light source (using the remote). The green light from the bulb and the reflected red light from the laser go into the optical chamber giving a wide green peak (corresponding to the bulb) and a sharp red peak (corresponding to the laser). Then we mark the point for the red peak. In the second step, we shine the green laser and turn the light bulb to red color and mark the sharp green peak for the second reference point of calibration. Using this method, we can quickly perform the calibration at any time without altering the setup or changing the source.

%% Adding AWG-conventional-similarity
   \begin{figure} [ht]
   \begin{center}
   \begin{tabular}{c} %% tabular useful for creating an array of images 
   \includegraphics[height=8cm]{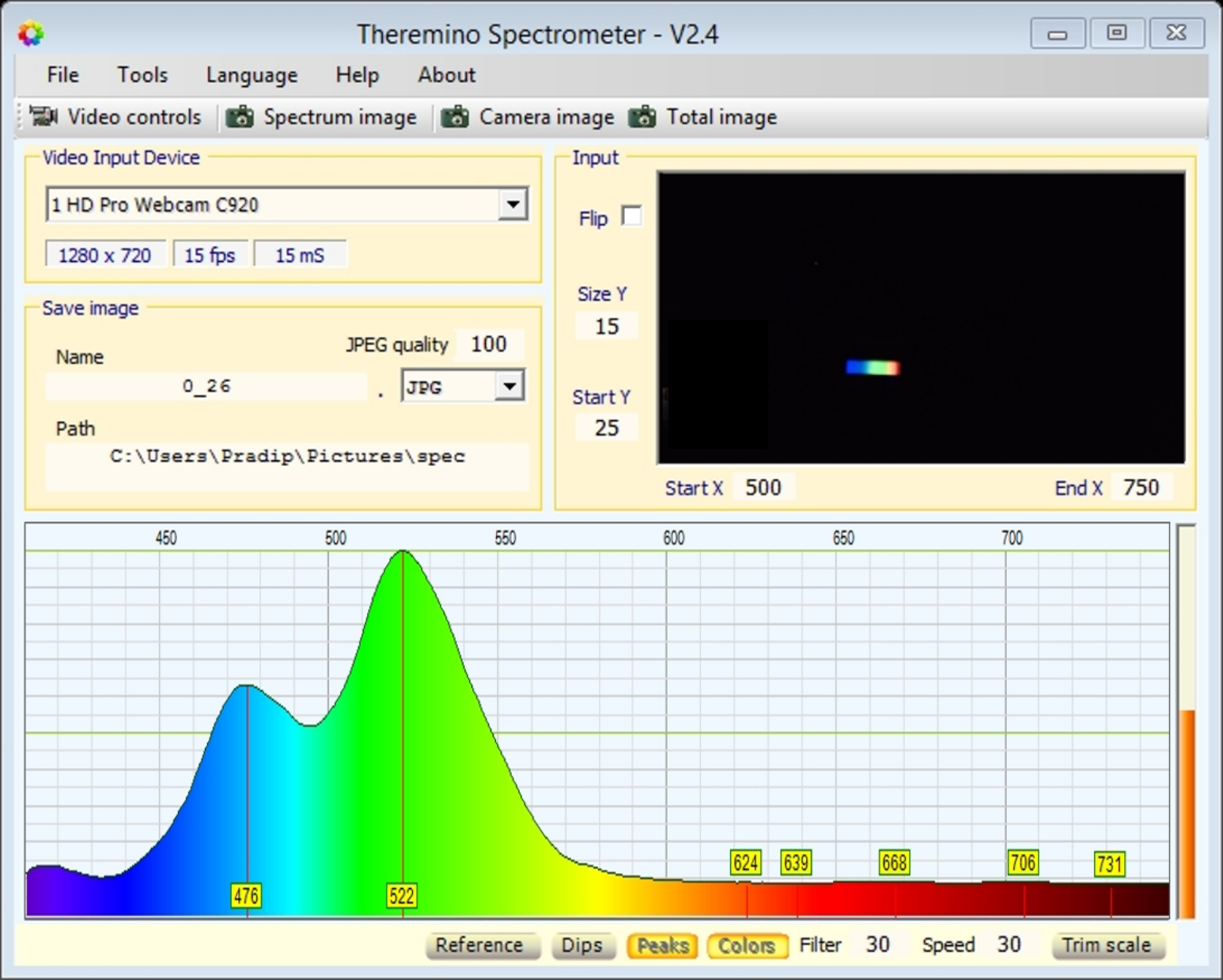}
   \end{tabular}
   \end{center}
   \caption[Picture of the spectrograph] 
%>>>> use \label inside caption to get Fig. number with \ref{}
   { \label{fig:AWG_sim_conventional} 
The interface of the theremino software \cite{Theremino}}
   \end{figure}    
   
   %% Adding AWG-conventional-similarity
   \begin{figure} [ht]
   \begin{center}
   \begin{tabular}{c} %% tabular useful for creating an array of images 
   \includegraphics[height=6cm]{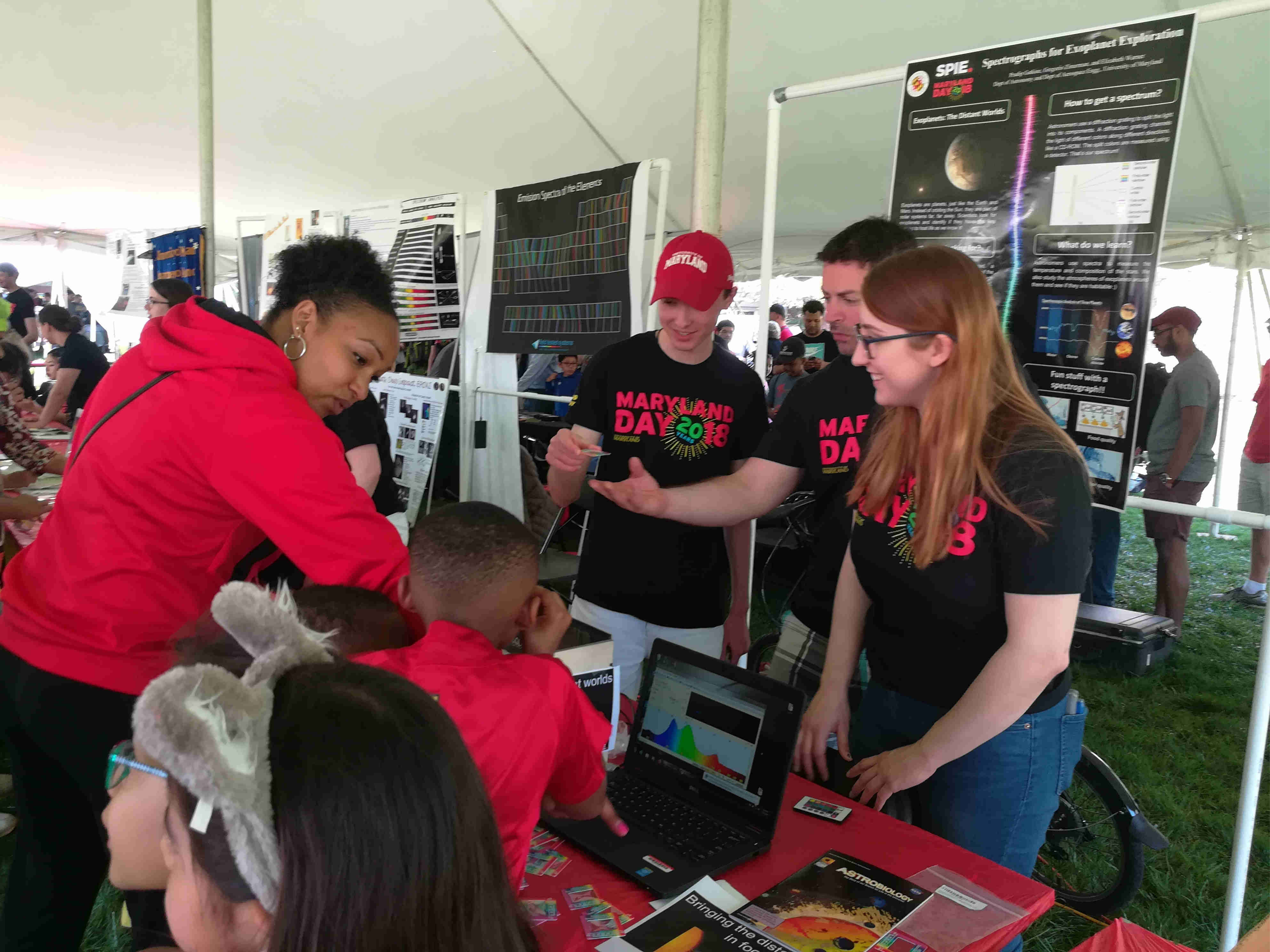}
   \end{tabular}
   \end{center}
   \caption[Picture of the spectrograph] 
%>>>> use \label inside caption to get Fig. number with \ref{}
   { \label{fig:AWG_sim_conventional} 
Volunteers demonstrating the spectrometer at Maryland Day 2018}
   \end{figure}

\section{Results}
The spectrometer kit presented here is a very versatile tool to demonstrate various concepts of optics and spectroscopy to an audience with a varying levels of optical experience, ranging from K-12 students/teachers to enthusiasts with minimal background in optics. This kit can be used in public events as well as in school demonstrations. We propose a non-exhaustive list of experiments that can be done with this DIY kit:

\noindent
1. Understanding the working principle of diffraction gratings,

\noindent
2. Understanding the role of coherence in spectroscopy by observing the effect of slit width on the spectrum,

\noindent
3. Studying absorption and emission spectra by using sources such as neon signs, elemental arcs ( such as hydrogen, helium, sodium, mercury, etc.) and various absorbing media,

\noindent
4. Studying water/food quality by using quantitative spectroscopic measurements,

\noindent
5. Studying the properties of polarization by using a polarized source or adding a polarizer in the optical assembly. This is especially helpful for experiments in organic chemistry to distinguish between enantiomers. 

\noindent
6. Studying the spectral signatures of milk, fruit juices, and various dyes used in food industry.\\ 

\noindent
{\bf Specific application case - Maryland Day:} 
Our first demonstration for a public audience was at Maryland Day in April 2018. Maryland day is UMD's largest and free community outreach event held annually in last week of April with 75000$+$ visitors, ranging from school students to teachers to parents. Our exhibit showcased interesting optical techniques to illustrate how optics is applied to look at distant and faint objects and how spectrometers help us understand chemical compositions of light sources, especially the exoplanets around stars. The visitors were able to play with these exhibits under the guidance and help by our volunteers to 'learn by doing'. 

During the period of the grant, we developed a spectrometer demonstration activity which can be used for outreach as well as classroom teaching. We had setup a booth at Maryland Day event at the University of Maryland, College Park. Our activity was visited by more than a thousand members the community throughout the day. Figure 7 shows a picture of our volunteers showing the spectrograph activity and engaging with a student and his teacher during the Maryland Day event. We believe, this activity has created its intended impact of spreading awareness and generating a sense of curiosity among general populous about how we gather knowledge about the universe. \\

\section{Summary and Future work}
With our exhibit at Maryland Day, we strove to educate our target audience about how the application of spectroscopy is making a revolution in the field of astronomy and exoplanets. We also involved undergraduates in building the exhibit and engaging with the audience while running the demonstration. In future, this exhibit is also slated to be used at the UMD observatory for demonstration to the visitors at the observatory's weekly open house, spreading the awareness among the neighborhood population.

\acknowledgments % equivalent to \section*{ACKNOWLEDGMENTS} 

The authors thank SPIE Education Outreach Grant for awarding a grant of \$3000 to the University of Maryland. We are thankful to the volunteers of Maryland Day who helped conduct the activity smoothly. We also thank Dr. Stuart Vogel from the Dept of Astronomy, University of Maryland and the UMD observatory for providing auxiliary support to develop the project and conduct the experiments.

% References
\bibliography{report} % bibliography data in report.bib
\bibliographystyle{spiebib} % makes bibtex use spiebib.bst

\end{document}